\documentclass[aps,10pt,prd,twocolumn,preprintnumbers,amsmath,amssymb,nofootinbib,superscriptaddress,a4paper,showpacs]{revtex4-1}

\usepackage[breaklinks]{hyperref}
\usepackage[english]{babel}
\usepackage[utf8]{inputenc}
\usepackage{graphicx}
\usepackage{color}
\usepackage{amsfonts,amsthm}
\usepackage{bm,bbm}

\newcommand{\be}{\begin{equation}}
\newcommand{\ee}{\end{equation}}
\newcommand{\ba}{\begin{eqnarray}}
\newcommand{\ea}{\end{eqnarray}}
\def\bs{\begin{subequations}}
\def\es{\end{subequations}}
\renewcommand{\leq}{\leqslant}
\renewcommand{\geq}{\geqslant}
\def\a{\alpha}
\def\b{\beta}
\def\de{\delta}

\def\la{\lambda}
\def\k{\kappa}

\def\om{\omega}

\def\vr{\varrho}

\def\cA{\mathcal{A}}

\def\cC{\mathcal{C}}

\def\cP{\mathcal{P}}

\def\cV{\mathcal{V}}

\def\ds{d_{\rm S}}
\def\dh{d_{\rm H}}

\def\deh{\de_{\rm H}}
\def\p{\partial}

\def\B{\Box}
\newcommand{\Eq}[1]{(\ref{#1})}
\def\com{\color{magenta}}
\def\cob{\color{blue}}
\newcommand{\book}[5]{\emph{#1} (#2, #3, #4, #5)}

\newcommand{\arX}[1]{\href{http://arxiv.org/abs/#1}{{\ttfamily\com arXiv:#1}}}
\newcommand{\doin}[6]{\href{http://dx.doi.org/#1}{{\cob #2 #3 {\bf #4}, #5 (#6)}}}
\newcommand{\doinn}[5]{\href{http://dx.doi.org/#1}{{\cob #2 {\bf #3}, #4 (#5)}}}
\newcommand{\doij}[5]{\href{http://dx.doi.org/#1}{{\cob #2 #3 (#5) #4}}}

\newcommand{\procsin}[5]{in \emph{#1}, edited by #2 (#3, #4, #5)} 

\newcommand{\tia}[1]{#1,}
\def\rme{e}
\def\rmd{d}
\def\rmi{i}
\newcounter{listcounter}

\begin{document}

\title{Multiscale spacetimes from first principles}

\author{Gianluca Calcagni}
\affiliation{Instituto de Estructura de la Materia, CSIC, Serrano 121, 28006 Madrid, Spain}

\date{September 7, 2016}

\begin{abstract}
Assuming only a smooth and slow change of spacetime dimensionality at large scales, we find, in a background- and model-independent way, the general profile of the Hausdorff and the spectral dimension of multiscale geometries such as those found in all known quantum gravities. Examples of various scenarios are given. In particular, we derive uniquely the multiscale measure with log oscillations of theories of multifractional geometry. Predictivity of this class of models and falsifiability of their abundant phenomenology are thus established.
\end{abstract}




\preprint{\doin{10.1103/PhysRevD.95.064057}{PHYSICAL REVIEW}{D}{95}{064057}{2017} \hspace{9cm} \arX{1609.02776}}

\maketitle


\section{Introduction and main result} 

The quest for a consistent theory of quantum gravitation unraveled the unexpected existence of generic properties of geometry found across all known frameworks \cite{tH93Car09}. First, that in a quantum setting geometry is anomalous and areas and volumes can behave very differently from their classical counterparts. Second, that the dimension of space or spacetime changes with the scale of observation. A relation between this \emph{dimensional flow} or \emph{multiscaling} and the ultraviolet (UV) properties of the theories was long since suspected but nowadays it has become clear that there is no direct universal link between the change of dimensionality and the microscopic finiteness of the quantum forces. However, it remains unclear why dimensional flow can often be described by similar asymptotic expressions in different theories. In this paper, we give an answer purely based on the infrared (IR) properties of the flow and independent of the dynamics, both for the Hausdorff dimension $\dh$ and for the spectral dimension $\ds$ of spacetime.

\newtheorem*{thm1}{First flow-equation theorem}
\begin{thm1}
Assume that
\begin{enumerate}
\item[(I)] dimensional flow of spacetime in dimension $d=\dh$ or $d=\ds$ is described by a continuous scale parameter $\ell$;
\item[(II)] this flow is slow at scales larger than a reference scale $\ell_*$ separating the IR from the UV; 
\item[(III)] effective spacetime is noncompact.
\end{enumerate} 
Then, at mesoscopic scales the most general real-valued spacetime dimension is
\be\label{dhdsgen}
d(\ell)\simeq D+b\left(\frac{\ell_*}{\ell}\right)^{c}+\text{\rm (log oscillations)},
\ee
where, to leading order, the log-oscillatory part is of the form $(\ell_2/\ell)^{c_1}\tilde F_\om(\ell)$, $c_1$ may differ from $c$, and the modulation factor $\tilde F_{\om}$ is given by
\be\label{tiF2}
\tilde F_{\om}\propto \cos\left[\om\ln\left(\frac{\ell}{\ell_\infty}\right)\right]\quad {\rm or}\quad \tilde F_{\om}\propto \sin\left[\om\ln\left(\frac{\ell}{\ell_\infty}\right)\right].
\ee
As a consequence, and up to an overall normalization, for $d=\dh$
\be\label{cvgen}
\cV(\ell) \simeq \ell^D\left[1-\frac{b_{\rm H}}{c_{\rm H}}\left(\frac{\ell_*}{\ell}\right)^{c_{\rm H}}\right]+\text{\rm (log oscillations)}
\ee
represents a generic Euclidean(ized) $D$-volume of linear size $\ell$. For $d=\ds$,
\be\label{cpgen}
\cP(\ell)\simeq \frac{1}{\ell^{D}}\left[1+\frac{b_{\rm S}}{c_{\rm S}}\left(\frac{\ell_*}{\ell}\right)^{c_{\rm S}}\right]+\text{\rm (log oscillations)}
\ee
is the return probability of spacetime. Exact expressions will be given in the text.
\end{thm1}

Terminology will be explained shortly. We will compare the flow in $\dh$ and $\ds$ of a number of popular theories of quantum gravity and find that they all realize the universal behavior predicted here. The sole and surprisingly simple reason is that dimensional flow is always slow in the IR and always reaches the topological dimension as an asymptote. This universal multiparametric form of the change in spacetime dimensionality must be complemented by information from the dynamics (which fixes the numerical values of the parameters) but it can have a great impact on model-independent phenomenology.

In parallel, we will also settle a long-standing issue concerning a class of theories of anomalous geometry known as multifractional spacetimes (see \cite{revmu} and references therein). These theories, three in total, had a double purpose originally. On one hand, to quantize gravity perturbatively in such a way that dimensional flow be under analytic control at all scales; this objective is still under pursue \cite{revmu}. On the other hand, by virtue of the supposed universality of certain properties of the flow, multifractional theories can be interpreted not only as stand-alone proposals, but also as effective descriptions of other independent multiscale theories of quantum gravity, most of which take considerable effort to produce usable phenomenology. The disarmingly easier way of multifractional theories to make contact with observations makes them an ideal testing ground of the type of phenomena we would expect if geometry was, as in all quantum gravities, multiscale. The measure being the same in all theories and their dynamics being relatively simple (except in the case with so-called fractional derivatives), it was possible to obtain a number of constraints from observations ranging from particle-physics and atomic scales to astrophysics and cosmology \cite{revmu}.

All theoretical and phenomenological results of multifractional theories rely on the assumption that the bare spacetime geometry (i.e., before considering the dynamics) is multifractal. In early papers, it was argued that dimensional flow can be well captured by a multifractal measure \cite{revmu} which, when coordinates factorize and the continuum approximation is taken, is of the form $\prod_\mu\rmd q^\mu:=\rmd q^0(x^0)\,\rmd q^1(x^1)\cdots\rmd q^{D-1}(x^{D-1})$ in $D$ integer topological dimensions, where each $q^\mu(x^\mu)$ is a real polynomial with noninteger exponents and logarithmic oscillations, decorated with a number of length scales $\ell_{1,2,3,\cdots}$. Omitting the label $\mu$ in the quantities $x^\mu$, $\a_\mu$, and $\ell_n^\mu$ below, the measure in each direction is
\ba
q(x) &=& x+\sum_{n=1}^{+\infty} \frac{\ell_n}{\a_n}\left|\frac{x}{\ell_n}\right|^{\a_n} F_n(x)\,,\label{meag}\\
F_n(x) &=& 1+A_n\cos\left(\om_n\ln\left|\frac{x}{\ell_\infty}\right|\right)+B_n\sin\left(\om_n\ln\left|\frac{x}{\ell_\infty}\right|\right)\!\!,\nonumber
\ea
where $\a_n\geq\a_{n+1},A_n,B_n$ are constants comprised between 0 and 1, $\om_n$ are frequencies, and 
\be\label{hier}
\ell_\infty\leq\cdots\leq\ell_n\leq\ell_{n-1}\leq\cdots\leq\ell_2\leq\ell_1\equiv\ell_*
\ee
are the fundamental scales of the geometry. $\ell_\infty$ can be identified with the Planck length \cite{revmu} and, for this reason, it is placed at the bottom of the hierarchy. Papers on the phenomenology of these models invoke the simplest multifractional example, the binomial measure
\bs\label{mea}\be
q_\a(x) = x+\frac{\ell_*}{\a}\left|\frac{x}{\ell_*}\right|^\a F_\om(x)\qquad \text{(for each direction)},
\ee
where
\be
F_\om(x)= 1+A\cos\left(\om\ln\left|\frac{x}{\ell_\infty}\right|\right)+B\sin\left(\om\ln\left|\frac{x}{\ell_\infty}\right|\right)\,.\label{Fom}
\ee\es
At scales $\ell\sim\ell_\infty$, the measure approximately enjoys the discrete scale invariance $q_\a(\la_\om x)\simeq \la_\om^\a q_\a(x)$ under the dilation $x\to\la_\om x$, where $\la_\om:=\exp(-2\pi/\om)$. At scales $\ell_\infty\ll\ell\sim\ell_*=\ell_1$, the log oscillations can be coarse grained and spacetime becomes effectively continuous, with Hausdorff dimension $\dh^{\rm UV}\simeq\sum_\mu\a_\mu$. At scales $\ell\gg\ell_*$, one recovers standard Minkowski spacetime with 
 $\dh\simeq D$. The binomial measure \Eq{mea} encodes the geometry at scales larger than the largest characteristic scale $\ell_1=\ell_*$, thus removing temporarily the need to consider the polynomial multifrequency measure \Eq{meag}. 

Despite its practical advantages, the whole multifractal spacetime paradigm has been criticized for relying on too strong a statement. Why using measures such as \Eq{meag} or \Eq{mea} and not others? If other measures realizing dimensional flow are possible, then are not multifractional theories \emph{ad hoc} and nonpredictive? We now state the answer, which is the second result of this paper:

\newtheorem*{thm2}{Second flow-equation theorem}
\begin{thm2}
Assume (I)--(III) as before and that
\begin{enumerate}
\item[(IV)] the measure is factorizable.
\end{enumerate}
Then,
\be\label{dmuf}
d=\sum_\mu d^{(\mu)}\,,
\ee
where for each direction $d^{(\mu)}$ is given by Eq.\ \Eq{dhdsgen} with $D=1$, $b\to b_\mu$, and $c\to c_\mu$. For $d=\dh$, the most general real measure is given by Eq.\ \Eq{meag} at all scales and by Eq.\ \Eq{mea} at mesoscopic scales. For a trivial flow, $\ell_n=0$ for all $n$.
\end{thm2}

If $c>0$, Eq.\ \Eq{dhdsgen} recovers the correct dimension at large scales, either from above ($b>0$) or from below ($b<0$). At intermediate scales, suitable choices of the parameters in Eq.\ \Eq{meag} reproduce any of the peaks and plateaux of $\ds$ found in quantum gravities. On the other hand, \Eq{meag} is an infinite perturbative expansion around the IR point and it may be inconvenient for writing analytic expressions of $d(\ell)\simeq\sum_\mu\a_{1,\mu}+O(\ell)$ in the deep UV. Incidentally, Eq.\ \Eq{dhdsgen} shows the existence of log oscillations in the mathematical definition of $\ds$, well known in examples of fractal geometry \cite{Akk12Akk2} but conjectured to be more general. In quantum gravity, these oscillations are either absent or often coarse grained with an averaging procedure, just like for the Hausdorff dimension \cite{revmu}. In general, they always appear when heat kernels and correlation functions are calculated on fractals. In multifractional theories, they are a long-range modulation of the geometry coming from the UV discrete scale invariance and they can leave a large-scale imprint on cosmological spectra as well as play a role in renormalization \cite{revmu}. Still, it is interesting to note that highly quantum states in loop quantum gravity (LQG), spin foams, and group field theory (GFT) can show a ``pathological'' (in the sense of very nonclassical) behavior such as a complex-valued $\ds$, where complex phases do not combine into real-valued oscillations \cite{revmu,COT1COT2COT3Thu15}.


\section{Assumptions}

Before proving the claims in italics, let us discuss the hypotheses. 
\begin{enumerate}
\item [(I)] A continuous parameter $\ell$ exists in all quantum gravities with a notion of distance, even when there is no fundamental notion of continuous spacetime. In continuous spacetimes, $\ell$ is an arbitrary length identified with the length scale at which we are probing the geometry, while in discrete and combinatorial settings (for instance, LQG and GFT) $\ell$ is measured in units of a lattice spacing or of the labels of combinatorial structures (e.g., complexes) \cite{COT1COT2COT3Thu15}. For the spectral dimension, the probing is often 
 imagined to take place via a diffusion process where $\ell$ plays the role of evolution parameter, involving a quantity $\cP(\ell)$ (possibly coming from the expectation value of operators on quantum states \cite{COT1COT2COT3Thu15}) called return probability. The interpretation of $\ell$ as a distance or as the inverse of resolution is possible even when geometry is fundamentally discrete and requires nonsmooth forms of calculi \cite{COT1COT2COT3Thu15}. Therefore, (I) is not restrictive and is fulfilled in all cases of interest to our knowledge. The Hausdorff dimension is defined as the scaling of the Euclideanized volume $\cV(\ell)$ of a $D$-ball of radius $\ell$ (or of a hypercube of edge $\ell$), while $\ds$ is the scaling of the return probability:
\bs\label{dhds}\ba
\dh(\ell)&:=&\frac{\rmd\ln\cV(\ell)}{\rmd\ln\ell}\,,\label{dhdsa}\\
\ds(\ell)&:=&-\frac{\rmd\ln\cP(\ell)}{\rmd\ln\ell}\,.
\ea\es
\item [(II)] Assumption (II) states that the profile $d(\ell)$ is flat near the IR endpoint $d^{\rm IR}={\rm const}$. This means that $d^{\rm IR}$ is reached as an asymptote at $\ell\to+\infty$, which is always the case if $\ell$ can be arbitrarily large. Then, $d^{\rm IR}=d(\infty)$. We have no counterexample entailing a ``maximum attainable length'' and (II) is general enough. 
\item [(III)] Hypothesis (III) is that spatial sections are noncompact, so that in physical situations the IR dimension of spacetime coincides with the topological dimension. Then, $d^{\rm IR}=d(\infty)=D$. In compact topologies, the $\ell\to+\infty$ limit corresponds to volumes wrapping around (or diffusion paths winding onto) space and the relation $d(\infty)=D$ is altered. A typical example is the 2-sphere, which is isomorphic to $\mathbb{R}^2$ only locally. In all calculations of $\dh$ and $\ds$, curvature effects must be ignored to prevent false positives \cite{Vas03}. 
\item [(IV)] Concerning the factorizability assumption in the second theorem, there have been attempts in the past to describe field theories on irregular geometries with nonfactorizable measures, but their range of applicability to physical situations was severely limited or nonextant \cite{Svo87Ey89aEy89bfra1fra2fra3}. The purely technical choice of defining multifractional theories with factorizable measures $\prod_\mu\rmd q^\mu$ has been successful in extracting observational constraints. Here, it will allow us to consider each spacetime direction separately, the total dimension of (Euclideanized) spacetime being Eq.\ \Eq{dmuf}. Of course, it may be that Nature, if multiscale, is not represented by factorizable geometries, in which case we have to look into other proposals. The first theorem will cover most of them.
\end{enumerate}
If spacetime is fundamentally continuous or embedded in a continuum (as in multifractional theories), then the first theorem applies exactly. If spacetime is continuous only after some coarse-graining, averaging, condensation, or semiclassical procedure (as in LQG/GFT), then Eq.\ \Eq{meag} is a good description of geometry only if the scale $\ell_\infty$ is larger than the UV cutoff $\ell_\textsc{uv}$ of the effective theory.


\section{Proof}

The proof of the flow-equation theorems uses only the properties of dimensional flow listed above; it does not depend on the dynamics of the theory. According to Eqs.\ \Eq{dhds}, the most natural parametrization of dimensional flow is via logarithmic scales. Therefore, it is convenient to employ the variable
\be
y:=\ln\ell
\ee
and the constants $y_n:=\ln \ell_n$, $y_*:=y_1$, and $y_\infty:=\ln\ell_\infty$, corresponding to the characteristic scales of the geometry. Integrating Eqs.\ \Eq{dhds}, we have
\bs\label{cvcp}\ba
\cV(y)&\propto&\exp\left[\int\rmd y\,\dh(y)\right]\,,\\
\cP(y)&\propto&\exp\left[-\int\rmd y\,\ds(y)\right].
\ea\es
Given that we will find the same formal expression for $\dh$ and $\ds$, dependent on a set of parameters $\la_j=c_j,b_j,\om_j,\dots$, at a formal level we can relate the $D$-volume $\cV(y,\la_j^{\rm H})$ and the return probability $\cP(y,\la_j^{\rm S})$ by
\be\label{form}
\cP(y,\la_j^{\rm S})=\frac{1}{\cV(y,\la_j^{\rm S})}\,.
\ee
Since $\la_j^{\rm H}\neq \la_j^{\rm S}$ in general, in any concrete theory one cannot say that the return probability is simply ``the inverse of the volume'' $\cV(y,\la_j^{\rm H})$. However, in this section we are interested in the functional form of these expressions, independently of the dynamically determined values of the parameters $\la_j^{\rm H}$ and $\la_j^{\rm S}$. Therefore, for the purpose of the proof Eq.\ \Eq{form} is a valid tool.

We also introduce two useful quantities: the difference
\be
\de_n(y):=d^{(n)}(y)-d^{(n-1)}(y)
\ee
(with $d^{(-1)}:=d^{\rm IR}$) of the dimension $d=\dh,\ds$ calculated at adjacent orders in an expansion we will introduce shortly, and the difference
\be
\de^{(n)}(y):=d^{(n)}(y)-d^{\rm IR}
\ee
between $d$ at order $n$ and the IR value.

At zero order, $\de_0=\de^{(0)}$. The case of trivial flow corresponds to the simple equation 
\be\label{fleq0}
\de_0=0\,,
\ee
i.e., $d^{(0)}=d^{\rm IR}$. This is the first example of relations between the dimension $d(\ell)$, its variation with respect to the scale $\ell$, and its IR value. As we will see, these relations are organized as an order expansion of the linear \emph{flow equation} with derivative order $n$
\be\label{flow}
\sum_{j=0}^n c_j\p_y^j\de_n=0\,,
\ee
where the $c_j$ are constants. The solution $\de_n$ is labelled by $n$. We do not see any immediate justification for generalizing this equation to nonlinear terms or nonconstant coefficients.

The rest of the section consists in solving the flow equation \Eq{flow} order by order.


\subsection{First theorem}

In $D$ dimensions for the general case, $d^{\rm IR}=D$ is the topological dimension, which coincides with the dimension at large scales. At $n=0$ order, integrating Eq.\ \Eq{fleq0} one gets $d=d^{(0)}=D$. For $d=\dh$, this corresponds to $\cV\propto\exp(Dy)=\ell^D$, ordinary Euclidean space with Lebesgue measure $\rmd\vr(x)=\rmd^Dx$. For $d=\ds$, one gets the return probability $\cP\propto \ell^{-D}$:
\ba
\cV&\propto&\ell^D\,,\qquad d=\dh=D\,,\qquad n=0\,,\label{cv0}\\
\cP&\propto&\frac{1}{\ell^D}\,,\qquad d=\ds=D\,,\qquad n=0\,.\label{cp0}
\ea

The next order brings information about the derivative of $d$ with respect to $y$. By virtue of assumption (II), $d'=\p_y d$ is approximately zero at sufficiently large scales, which means that the profile $d(y)$ is almost flat. To get a nontrivial description of geometry, we must combine the information coming from $d^{(0)}\simeq d^{\rm IR}$ with that from $d'\simeq 0$, in such a way that the dimension and its first derivative are nonzero separately with increasingly good approximation. This is achieved by the first-order flow equation [$n=1$ and $c_1=1$ in Eq.\ \Eq{flow}]
\be
\de_1'+c_0\de_1=0\,.
\ee
Integrating and noting that $\de^{(1)}=\de_1$, we find
\be\label{deone}
\de^{(1)}(y)=\de_1(y)=b\exp\left[-c_0(y-y_*)\right]\,,
\ee
where $b$ and $y_*$ are arbitrary constants. This expression reproduces the second term in Eqs.\ \Eq{dhdsgen}, \Eq{cvgen}, and \Eq{cpgen}, where $c_0=c$ and $\ell_*=\rme^{y_*}$ is the first fundamental scale of the geometry encountered when running from the IR. In fact, $d(\ell)=d^{\rm IR}+\de^{(1)}$, so that
\bs\label{dhdsgen1}\ba
{\dh}^{(1)}&=& D+b_{\rm H}\left(\frac{\ell_*}{\ell}\right)^{c_{\rm H}},\\
{\ds}^{(1)}&=& D+b_{\rm S}\left(\frac{\ell_*}{\ell}\right)^{c_{\rm S}},
\ea\es
while from Eqs.\ \Eq{cvcp} and \Eq{deone} one has
\bs\ba
\cV &\propto& \exp\int\rmd y\,[D+\de_1(y)]\nonumber\\
    &=&\exp\left\{Dy-\frac{b_{\rm H}}{c_{\rm H}}\exp\left[-c_{\rm H}(y-y_*)\right]\right\}\nonumber\\
    &=&\ell^D\exp\left[-\frac{b_{\rm H}}{c_{\rm H}}\left(\frac{\ell_*}{\ell}\right)^{c_{\rm H}}\right]\label{cvcp2a}\\
		&\simeq& \ell^D\left[1-\frac{b_{\rm H}}{c_{\rm H}}\left(\frac{\ell_*}{\ell}\right)^{c_{\rm H}}\right]\,,\label{cvcp2b}\\
\cP &\propto& \frac{1}{\ell^D}\exp\left[\frac{b_{\rm S}}{c_{\rm S}}\left(\frac{\ell_*}{\ell}\right)^{c_{\rm S}}\right]\label{cvcp2c}\\
		&\simeq& \frac{1}{\ell^D}\left[1+\frac{b_{\rm S}}{c_{\rm S}}\left(\frac{\ell_*}{\ell}\right)^{c_{\rm S}}\right]\,.\label{cvcp2d}
\ea\es
In general, $b_{\rm H}\neq b_{\rm S}$ and $c_{\rm H}\neq c_{\rm S}$. These coefficients are determined by the dynamics of the model.

At $n=2$, the general solution of the flow equation (here the factor of $2$ is for convenience)
\be
\de_2''+2c_1\de_2'+c_0\de_2=0
\ee
is $\de_2(y)=\rme^{-c_1(y-y_2)}[b_+\,\exp(-\sqrt{c_1^2-c_0}y)+b_-\,\exp(\sqrt{c_1^2-c_0}y)]$. The dimension is
\be
d^{(2)}=d^{(1)}+\de_2=d^{\rm IR}+\de_1+\de_2\,.
\ee
The case $c_1^2-c_0\geq 0$ can be ignored without loss of generality because pure polynomial profiles are recovered by the most general higher-order solutions [here, we have either the first-order solution ($c=c_1=\sqrt{c_0}$) or (exactly or in the mesoscopic/IR limit $y\gg 1$) a trinomial profile $d^{(2)}\propto D+b(\ell/\ell_1)^{-Dc}+\tilde b(\ell/\ell_2)^{-Dc_1}$]. Setting thus $\om^2:=c_0-c_1^2>0$ and keeping the symbol $c$ for the first-order coefficient, we get the expression of $\de^{(2)}=d^{(2)}-d^{\rm IR}=\de_1+\de_2$:
\ba
\hspace{-0.5cm}\de^{(2)}(y) &=& b\,\rme^{-c(y-y_*)}\nonumber\\
&&+\rme^{-c_1(y-y_2)}\left(b_+\,\rme^{-\rmi\om y}+b_-\,\rme^{\rmi\om y}\right)\!.\label{de2}
\ea
This is the most general second-order solution, with $c_1\geq c$ (for consistency, the $n=2$ correction cannot dominate over the $n=1$ one) and complex roots. The dimension where running occurs (Hausdorff and/or spectral) is
\be
d^{(2)}=D+b\left(\frac{\ell_*}{\ell}\right)^c+\left(\frac{\ell_2}{\ell}\right)^{c_1}\left(b_+\,\ell^{-\rmi\om}+b_-\,\ell^{\rmi\om}\right),
\ee
where the set $b=b_{\rm H}$, $c=c_{\rm H}$, $c_1=c_{1,{\rm H}}$, $\b_\pm=\b_{\pm,{\rm H}}$, and $\om=\om_{\rm H}$ for $d=\dh$ may be different from the parameters $b=b_{\rm S}$, $c=c_{\rm S}$, $c_1=c_{1,{\rm S}}$, $\b_\pm=\b_{\pm,{\rm S}}$, and $\om=\om_{\rm S}$ for $d=\ds$. The volume and return probability read
\bs\ba
\cV &\propto& \ell^D\exp\left[-\frac{b_{\rm H}}{c_{\rm H}}\left(\frac{\ell_*}{\ell}\right)^{c_{\rm H}}-\left(\frac{\ell_2}{\ell}\right)^{c_{1,{\rm H}}}\tilde F_{\rm H}(\ell)\right]\label{cvcp5a}\\
&\simeq& \ell^D\left[1-\frac{b_{\rm H}}{c_{\rm H}}\left(\frac{\ell_*}{\ell}\right)^{c_{\rm H}}-\left(\frac{\ell_2}{\ell}\right)^{c_{1,{\rm H}}}\tilde F_{\rm H}(\ell)\right],\label{cvcp5b}\\
\cP &\propto& \frac{1}{\ell^D}\exp\left[\frac{b_{\rm S}}{c_{\rm S}}\left(\frac{\ell_*}{\ell}\right)^{c_{\rm S}}+\left(\frac{\ell_2}{\ell}\right)^{c_{1,{\rm S}}}\tilde F_{\rm S}(\ell)\right]\label{cvcp5c}\\
&\simeq& \frac{1}{\ell^D}\left[1+\frac{b_{\rm S}}{c_{\rm S}}\left(\frac{\ell_*}{\ell}\right)^{c_{\rm S}}+\left(\frac{\ell_2}{\ell}\right)^{c_{1,{\rm S}}}\tilde F_{\rm S}(\ell)\right],\label{cvcp5d}
\ea\es
where
\bs\label{tiFi}\ba
\tilde F_{\rm H}(\ell) &=& \tilde b_{+,{\rm H}}\,\ell^{-\rmi\om_{\rm H}}+\tilde b_{-,{\rm H}}\,\ell^{\rmi\om_{\rm H}},\\
\tilde F_{\rm S}(\ell) &=& \tilde b_{+,{\rm S}}\,\ell^{-\rmi\om_{\rm S}}+\tilde b_{-,{\rm S}}\,\ell^{\rmi\om_{\rm S}},
\ea\es
and $\tilde b_\pm=b_\pm/(c_1\pm\rmi\om)$. If $\tilde b_+= \tilde b_-^*\propto \rme^{\rmi\om y_\infty}$, then one obtains the real-valued logarithmic oscillations \Eq{tiF2} in Eqs.~\Eq{dhdsgen}, \Eq{cvgen}, and \Eq{cpgen}.


\subsection{Second theorem}

For multifractional theories, the length $\cV^{(\mu)}$ along the $\mu$th direction in a continuous space is just the integral $\cV^{(\mu)}(\ell)=\int_{\ell_\textsc{uv}}^\ell\rmd\vr(x^\mu)=\vr(\ell)-\vr(\ell_\textsc{uv})$ of the spacetime measure $\vr(x^\mu)$ from a UV cutoff $\ell_\textsc{uv}$; the constant $\vr(\ell_\textsc{uv})$ does not affect Eq.\ \Eq{dhdsa} and is zero in these theories. The volume of a hypercube is simply $\cV\propto\prod_\mu\cV^{(\mu)}$ (the $D$-volumes of other objects differ in their normalization but not in the general scaling), while the return probability is the product of $D$ profiles: $\cP\propto\prod_\mu\cP^{(\mu)}$. Therefore, the dimension of spacetime is the sum of the dimensions along each direction, Eq.\ \Eq{dmuf} with $d^{{\rm IR},\mu}=1$, and Eqs.\ \Eq{cvcp} remain valid.

An expression we will need is the approximate profile $\dh(y)$ from Eq.\ \Eq{mea} at mesoscopic scales. Working in the positive half-line $x>0$ for simplicity, and omitting labels $\mu$ everywhere, for $\vr=q_\a$ we have 
\ba
\deh(y)&:=&\dh(y)-\dh^{\rm IR} \simeq  -(\a^{-1}-1)\rme^{(\a-1)(y-y_*)}F_\om(y)\,,\nonumber\\
F_\om(y)
&=&1+\cA_-\,\rme^{\rmi\om(y-y_\infty)}+\cA_+\,\rme^{-\rmi\om(y-y_\infty)},\label{dhint2}
\ea
where $\cA_\pm=(A\pm\rmi B)[1\pm\rmi\om/(1-\a)]/2$ (notice that $\deh$ is real-valued).

At $n=0$ order, the results \Eq{cv0} and \Eq{cp0} are recovered. For $n=1$, to reproduce the dimensional flow of $\dh$ in multifractional theories without log oscillations, we must compare Eq.\ \Eq{deone} with Eq.\ \Eq{dhint2} when $F_\om=1$: then, 
\be\label{fix1}
b_{{\rm H},\mu}=-(\a_\mu^{-1}-1)\,,\qquad c_{{\rm H},\mu}=1-\a_\mu\,.
\ee
This is just a redefinition of labels, since the constants $y_*$ and $c_{\rm H}$ are mutually independent just like $\ell_*$ and $\a$. Signs are also unconstrained: we can set $\a<1$ ($b_{\rm H}<0$, $c_{\rm H}>0$) only in the case we want a dimensional flow where $\dh^{\rm UV}<\dh^{\rm IR}$. This requirement may be desirable for phenomenology but it plays no role here. The coefficients in the spectral dimension are more model-dependent. For instance, in the theory with weighted derivatives $\ds=D$ is constant ($b_{{\rm S},\mu}=0$), while in the theory with $q$-derivatives $\ds\simeq\dh$ at all the plateaux of dimensional flow \cite{revmu}.

Integrating the solution \Eq{deone} and expanding in $\ell\gg\ell_*$ (consistently with the mesoscopic-scale approximation entailed in the $n=1$ truncation), we can get $\cV$ and $\cP$ from Eq.\ \Eq{cvcp}. Taking into account factorizability, one has
\bs\label{dhdsgen1muf}\ba
{\dh}^{(1)}&=& D+\sum_{\mu=0}^{D-1}b_{{\rm H},\mu}\left(\frac{\ell_*}{\ell}\right)^{c_{{\rm H},\mu}},\\
{\ds}^{(1)}&=& D+\sum_{\mu=0}^{D-1}b_{{\rm S},\mu}\left(\frac{\ell_*}{\ell}\right)^{c_{{\rm S},\mu}},
\ea\es
and
\bs\label{cvcp3}\ba
\cV &\propto& \ell^D\exp\left[-\sum_{\mu=0}^{D-1}\frac{b_{{\rm H},\mu}}{c_{{\rm H},\mu}}\left(\frac{\ell_*^\mu}{\ell}\right)^{c_{{\rm H},\mu}}\right]\label{cvcp3a}\\
		&\simeq& \ell^D\left[1-\sum_{\mu=0}^{D-1}\frac{b_{{\rm H},\mu}}{c_{{\rm H},\mu}}\left(\frac{\ell_*^\mu}{\ell}\right)^{c_{{\rm H},\mu}}\right]\,,\label{cvcp3b}\\
\cP &\propto& \frac{1}{\ell^D}\exp\left[\sum_{\mu=0}^{D-1}\frac{b_{{\rm S},\mu}}{c_{{\rm S},\mu}}\left(\frac{\ell_*^\mu}{\ell}\right)^{c_{{\rm S},\mu}}\right]\label{cvcp3c}\\
		&\simeq& \frac{1}{\ell^D}\left[1+\sum_{\mu=0}^{D-1}\frac{b_{{\rm S},\mu}}{c_{{\rm S},\mu}}\left(\frac{\ell_*^\mu}{\ell}\right)^{c_{{\rm S},\mu}}\right]\,.\label{cvcp3d}
\ea\es
Notice that the scale $\ell$ is always taken to be the same along all directions. When geometry is time-space isotropic, these expressions simplify to Eqs.\ \Eq{cvcp2a}--\Eq{cvcp2d} with 
\be
b_{\rm H}\to Db_{\rm H}\,,\qquad b_{\rm S}\to Db_{\rm S}\,.
\ee
This shows that the second flow-equation theorem is little more than a corollary of the first.

We have just obtained, for $d=\dh$, the measure $\vr(x)\simeq x+(\ell_*/\a)({x}/{\ell_*})^\a$ in the absence of log oscillations. The binomial measure \Eq{mea} with $F_\om=1$ is the approximation of the full log-oscillating measure at scales above $\ell_\infty$. Therefore, if the description in terms of flow equations is correct and self-consistent, there is a very natural way in which we can obtain log oscillations: to consider higher-order versions of \Eq{flow}. In fact, it is necessary to go to second order, $n=2$, as we saw in the previous subsection. We do not repeat the calculation here: at the end of the day, Eqs.\ \Eq{dhdsgen1muf} and \Eq{cvcp3} are augmented by a factor $(\ell_2/\ell)^{c_1}\tilde F_{\om_\mu}(\ell)$ in the sums over $\mu$, where
\be\label{tiF}
\tilde F_{\om_\mu}(\ell)=\tilde b_{+,\mu}\,\ell^{-\rmi\om_{\mu}}+\tilde b_{-,\mu}\,\ell^{\rmi\om_{\mu}}.
\ee
Again, in the fully isotropic configuration $\om_\mu=\om$, $\tilde b_{\pm,\mu}=\tilde b_\pm$, and when $\tilde b_+\propto \tilde b_-^*\propto \rme^{\rmi\om y_\infty}$, one has the modulation factors \Eq{tiF2}. Going to third and fourth order, matching the parameters order to order (exponents, frequencies, and so on) and taking the fully isotropic configuration, one can easily obtain the modulation factor \Eq{Fom}. Comparing with Eq.\ \Eq{dhint2}, the binomial measure \Eq{mea} corresponds to $b$ and $c$ reparametrized as in Eq.\ \Eq{fix1} and with $b_\pm=-(\a^{-1}-1)\cA_\pm\rme^{(1-\a)y_*\pm\rmi\om y_\infty}$, $c_0=\om^2+(1-\a)^2$, and $c_1=1-\a$. There is one independent constant less, since $c_1=c$, but this choice maximizes the chance to get nontrivial effects from log oscillations at scales $\sim\ell_*$. Therefore, Eq.\ \Eq{mea} is the most general $n=4$ real-valued factorizable solution of the flow equations for $\dh$ such that large-scale effects of log oscillations are maximized at scales $\sim\ell_*$. 

Note, however, that the effect of log oscillations can be maximized even in the IR by setting $c_1=0$. Moreover, the most general fully isotropic case where different-order parameters are not matched is Eq.~\Eq{tiF}, which has log-average $\langle \tilde F_\om\rangle=0$, contrary to the modulation factor \Eq{Fom} typically used in multifractional theories and such that $\langle F_\om\rangle=1$. This fact could play a very important role in reinterpreting the microscopic structure of these spacetimes \cite{revmu}.


\subsection{All orders}

The method of the flow equation starts from the IR and hits first the largest scale $\ell_1=\ell_*$, and then the lowest scale $\ell_\infty$ in the hierarchy \Eq{hier} of the multiscale paradigm. There is no contradiction in having found the largest and the smallest scale of the hierarchy at $n=2$. Log oscillations are an independent structure with respect to the polynomial behavior of the measure \cite{revmu} and they modulate the geometry even at scales much larger than $\ell_\infty$. Scales below $\ell_*$ are too small to be constrained by experiments, which can only say something about $\ell_*$. The latter ``screens'' the microscopic structure of the measure but it does not prevent the logarithmic modulation to manifest itself in subtle ways \cite{revmu}. Then, at $n=2$ the derivative expansion has both the expected range in the dimensional flow ($\sim\ell_*$) and enough sensitivity to catch the modulation structure. 

The second-order solution is an approximation of more complicated measures. The general solution of the $n$th-order flow equation \Eq{flow} is a complex superposition
\be\label{soln}
\de_n=\sum_{i=0}^{n-1} b_{i,n}\,\exp(k_{i,n}y)\,,\qquad \sum_{j=0}^n c_j k_{i,n}^j=0
\ee
of exponentials with $n$ complex wavenumbers $k_{i,n}$ satisfying a characteristic equation for all $i$. Physical solutions do not have exactly $n$ distinct roots because they should be (and can always be, by choosing the $c_j$ and $k_i$) real-valued (a condition lifted in Ref.\ \cite{revmu}), positive semidefinite, and with the correct IR asymptote. Getting thus new scales, frequencies, and amplitudes in the dimension $d=\sum_{n=0}^{+\infty}\de_n$, we enter the realm of the polynomial multimodal measure \Eq{meag} for $d=\dh$, or of a polynomial multimodal return probability for $d=\ds$.

\section{Comparison with quantum gravities} 

We conclude by comparing Eq.\ \Eq{dhdsgen} with several theories of quantum gravity. Asymptotic safety \cite{LaR5RSnaxCES}, causal dynamical triangulations (CDT) \cite{AJL4BeHSVW1}, spacetimes near black holes \cite{CaGMur12,ArCa1}, nonlocal gravity and string field theory \cite{Mod11BGKM,CaMo1} all have trivial dimensional flow in the Hausdorff dimension ($\dh=D$). Noncommutative spacetimes usually have $\dh=D$ \cite{Ben08AAArTr1}, but in the case of $\kappa$-Minkowski with cyclic-invariant action $b<0$ and $c=1$ \cite{revmu}. Finally, states of LQG and GFT describing general discrete quantum geometries display the kink profile of the binomial measure \Eq{mea} without log oscillations \cite{COT1COT2COT3Thu15}. In the analytic example of the lattice $\cC_\infty=\mathbb{Z}^{D-1}$, the Hausdorff dimension reads $\dh=2+O(\ell)$ in the UV ($\ell$ is measured in units of the lattice spacing), while in the IR $\dh=D-(D-1)(D-2)/(2\ell)+O(\ell^{-2})$, giving $b<0$ and $c=1$. 

Concerning the spectral dimension near the IR, the profile \Eq{dhdsgen} reproduces the one in the multifractional theories with weighted and $q$-derivatives \cite{revmu}. In particular, in the theory with $q$-derivatives $b=\a-1$ and $c=1-\a$ for each direction. The log-oscillatory modulation is also recovered. In asymptotic safety, $\ell$ is the IR cutoff governing the renormalization-group equation of the metric \cite{LaR5RSnaxCES}. The multiscale profile of the spectral dimension is calculated analytically at each plateau and numerically in transition regions. The author is unaware of any semianalytic approximation giving $b$ and $c$ in \Eq{dhdsgen}. The same holds for Ho\v{r}ava--Lifshitz gravity. The rest of the models listed from now on have $c=2$, without exception. In CDT, $b<0$ is found numerically \cite{AJL4BeHSVW1}. In a nonlocal field-theory model near a black hole, $b=(D+1)/2$ \cite{ArCa1}. In fuzzy spacetimes, $b=-D$ \cite{MoN}. In nonlocal gravity with $\rme^\B$ operators as in string field theory, $b<0$ (one can show that $b=-36$ in $D=4$) \cite{CaMo1}. The noncommutative examples of \cite{Ben08AAArTr1} are the following: in $D=3$ Einstein gravity with quantized relativistic particles, $b=-21/16$; in Euclidean $\k$-Minkowski space with bicovariant Laplacian and AN(3) momentum group manifold, $D=4$ and $b=-2$; with AN(2) momentum group manifold, $D=3$ and $b=-3/2$; with bicrossproduct Laplacian, $D=4$ and $b=1$. In LQG and GFT, one can check numerically that $b>0$ for all the classes of states inspected \cite{COT1COT2COT3Thu15}. In all cases, $c=2$. 

More examples can be found in Ref.\ \cite{revmu}. In this companion paper, we will also discuss in greater detail some of the physical implications of the flow-equation theorems, including the resolution of the so-called presentation problem, a major contribution towards a definition of the theory with fractional derivatives, and consequences for the priors on the length hierarchy, complex dimensions, the big-bang singularity, and renormalization.


\section*{Acknowledgments} 

The author is under a Ram\'on y Cajal contract and is supported by the I+D grant FIS2014-54800-C2-2-P.


\end{document}